\documentstyle[a4,11pt]{article}
\begin{document}  
\begin{center}
Notes on Riemann geometry and integrable systems. Part V.
\end{center}

\begin{center}
Kur.Myrzakul\footnote{Permanent address: Institute of Mathematics, 
Alma-Ata, Kazakhstan} and R.Myrzakulov\footnote{Permanent 
address: Institute of Physics and Technology,
480082,  Alma-Ata-82,  Kazakhstan. E-mail: cnlpmyra@satsun.sci.kz}
\end{center}

\begin{center}
Department of Mathematics, Bilkent University, 06533 Ankara, Turkey 
\end{center}

\begin{abstract}
Some aspects of multidimensional soliton geometry are 
considered. 
\end{abstract}

\tableofcontents
\section{Introduction}
The purpose of this note is to consider the relationship of some 
multidimensional spaces and some partial differential equations (PDE).
The study of the deep relationship between differential geometry and 
PDE has a long and distinguished history, 
going back to the works of Darboux, Lie, Backlund, Goursat and E.Cartan. 
This relationship stems from the fact that most of the local properties 
of manifolds are expressed naturally in terms of PDE [1-25]. 

There are several reasons for our interest in the study of the 
relationship between Riemann space and integrable PDE in 2+1 dimensions
(see, e.g., [26-30]). 
One reason has been motivated by revealed connections between 
corresponding problems of classical differential geometry and modern 
problems of mathematical and theoretical physics. 
A more specific reason for interest is that the (1+1)-dimensional 
counterparts of the known integrable PDE in 
2+1 dimensions, such as the Kadomtsev-Petviashvili (KP) equation, its 
some modifications, the Davey-Stewartson (DS) equation, the 
three-dimensional three-wave resonant interaction system, and others, 
have been shown to have natural geometrical interpretations [1-24]. 

The classical differential geometry serves as an injector of many 
equations integrable in this or that sense [1-30]. Among them, integrable 
spin 
systems and their relativictic counterparts - sigma models  play especially 
remarkable role [32, 39, 50-53].  It turns out that one of the central 
problems of 
the differential geometry: the problem of constructing  $n$ - 
curvilinear coordinate systems is deeply connected to the theory of 
integrable spin systems and sigma models [54-57]. 

This note is a sequel to the 
preceding notes [54-57]. 
The layout out of this paper is as follows. In section 2 we consider the 
three-dimensional flat space obtaining the main geometrical equations. 
Next, in section 3 we combine these equations in the $n=3$ case with the 
several multidimensional integrable spin systems (MISS) to get integrable 
reductions of the 3-nonorthogonal coordinate systems (NOCS). 
In section 4, we conjecture that the 4-NOCS also admit integrable 
reduction which is connected with the Self-Dual Yang-Mills 
(SDYM) equation. Finally, in section 5 some integrable aspects 
of soliton immersions in multidimensions are considered.

\section{Flat 
Riemann  space}
Let $V^{n}$ be the space endowed with the affine connection. In this 
space we introduce two systems of coordinates: $(x)=(x^{1}, x^{2}, 
... , x^{n})$ and $(y)=(y^{1}, y^{2}, ... ,  y^{n})$. It is well known from 
the classical differential geometry that these coordinate systems are 
connected by the following set of equations of second order (remark: 
for convenience, in [54-57] and this note we will use the unified 
common numerations for formulas) 
$$ \frac{\partial^{2}y^{k}}{\partial 
x^{i}\partial 
x^{j}}= \Gamma^{l}_{ij}(x)\frac{\partial y^{k}}{\partial x^{l}}-
\Gamma^{k}_{lm}(y)\frac{\partial y^{l} \partial y^{m}}{\partial 
x^{i}\partial x^{j}}.   \eqno(228)
$$
In this paper we assume Einstein summation convention, i.e., the repeated 
indices are summed up.  The curvature tensor has the form
$$
R^{i}_{klm}=[\frac{\partial \Gamma_{m}}{\partial x^{l}}
-\frac{\partial \Gamma_{l}}{\partial x^{n}}
+\Gamma_{l}\Gamma_{m}-
\Gamma_{m}\Gamma_{l}]^{i}_{k}.
\eqno(229)
$$           
Let the space $V^{n}$ is flat and let   the metric tensor 
in $V^{n}$ in the 
coordinate system $(y)$ is diagonal:
$$
ds^{2}=\mu_{ij}dy^{i}dy^{j}   \eqno(230)
$$
with $\mu_{ij}=\pm 1$. Hence follows that  the corresponding connection 
and   Riemann's curvature tensor are equal to zero: 
$$
\Gamma^{i}_{jk}(y)=0,  \quad R^{i}_{klm}(y)=0.
\eqno(231)
$$

Let for the coordinate system $(x)$ the metric  has the form
$$
ds^{2}=g_{ij}dx_{i}dx_{j}.   \eqno(232)
$$                        
As the curvature tensor has the law of transformation as the four rank 
tensor
$$
R^{i}_{klm}(x)=\frac{\partial x^{i}}{\partial y^{s}}
\frac{\partial y^{r}}{\partial x^{k}}
\frac{\partial y^{q}}{\partial x^{l}} 
\frac{\partial y^{p}}{\partial x^{n}}
R^{i}_{klm}(y)
\eqno(233)
$$                                             
the curvature tensor for 
the coordinate system 
$(x)$ is also equal to 
zero
$$
R^{i}_{klm}(x)=0.  \eqno(234) 
$$                                                
Hence,  
for the coordinate system $(x)$ we get the following system of 
three equations
$$
\frac{\partial \Gamma_{m}}{\partial x^{l}}
-\frac{\partial \Gamma_{l}}{\partial x^{n}}
+\Gamma_{l}\Gamma_{m}-
\Gamma_{m}\Gamma_{l}=0
\eqno(235)
$$  
where $\Gamma_{m}(x)$  are matrices with components 
$$
\Gamma^{k}_{ij}=\frac{1}{2}g^{kl}(g_{il,j}+g_{jl,i}-g_{ij,l}). \eqno(236)
$$
Eqs. (235) are the compatibility condition of the following overdetermined 
system of the linear equations
$$
\Phi_{,k}=\Gamma_{k}\Phi. \eqno(237)
$$

We note that in our case the scalar 
curvature is equal to 
zero $$ R=\sum^{3}_{i,k,l,m=1}g^{il}g^{km}R_{iklm}=0.  \eqno(238)
$$
 
Now the system (228) takes the form
$$
\frac{\partial^{2}y^{k}}{\partial x^{i}\partial x^{j}}=
\Gamma^{l}_{ij}(x)\frac{\partial y^{k}}{\partial x^{l}}.  \eqno(239)
$$ 

Let ${\bf r}=(y^{1},y^{2}, ... , y^{n}) ={\bf r}(x^{1},x^{2}, ... 
, x^{n})$ is the position vector. 
Then as follows from (165) the position vector 
${\bf r}$ 
satisfies the following set of equations
$$
{\bf r}_{,ij} = \Gamma^{k}_{ij} {\bf r}_{,k}.    \eqno(240)
$$ 
We can rewrite the equation (240) in the 
following form 
$$ 
Z_{,k} = A_{k}Z \eqno(241) 
$$
where
$$
Z=({\bf r}_{,k})^{T}
\eqno(242)
$$ 
and
$$
A_{k} =
\left ( \begin{array}{cccc}
\Gamma^{1}_{k1} & \Gamma^{2}_{k1} & ... & \Gamma^{n}_{k1} \\
\Gamma^{1}_{k2} & \Gamma^{2}_{k2} & ... & \Gamma^{n}_{k2}\\
...               &   ...              & ... & ...               \\
\Gamma^{1}_{kn}            &\Gamma^{2}_{kn}&...&     \Gamma^{n}_{kn}
\end{array} \right).   
$$
The compatibility condition of these equations gives
$$
A_{i,j}-A_{j,i}+[A_{i},A_{j}]=0. \eqno(243)
$$
Now we introduce the triad unit vectors by the 
  way $$
{\bf e}_{1} = \frac{{\bf r}_{,1}}{H_{1}}, \quad    
{\bf e}_{l} = \frac{{\bf r}_{,l}}{H_{l}} +\sum^{n}_{k=1, k\neq 
l}c_{kl}{\bf r}_{,k}.
\eqno(244)
$$
Here
$$
H_{k}=\mid{\bf r}_{,k}\mid, \quad {\bf e}_{1}^{2}=\beta=\pm 1, \quad 
{\bf e}_{k\neq 1}^{2}=1.  \eqno(245) $$
The equations  for ${\bf e}_{k}$ take the form 
$$
\left ( \begin{array}{c}
{\bf e}_{1} \\
{\bf e}_{2} \\
... \\
{\bf e}_{n}
\end{array} \right)_{,k}= B_{k}
\left ( \begin{array}{c}
{\bf e}_{1} \\
{\bf e}_{2} \\
...\\
{\bf e}_{n}
\end{array} \right)
\eqno(246)
$$
with
$$
B_{k} = (b_{ij}^{k}), \quad B_{k}\in so(p,q), \quad p+q=n. \eqno(247)
$$
The integrability condition of the system is
$$
\frac{\partial B_{i}}{\partial x^{j}}-\frac{\partial B_{j}}{\partial
x^{i}}  + [B_{i},B_{j}] = 0.  \eqno(248)
$$
Now we want consider some particular cases, namely, the cases $n=3$ and 
$n=4$ and present some integrable reductions in these cases.
 
\section{The n=3 case}

So we consider the case $n=3$. In this case the unit vectors ${\bf 
e}_{k}$ we define as 
$$
{\bf e}_{1} = \frac{{\bf r}_{x}}{H_{1}}, \quad
{\bf e}_{2} = \frac{{\bf r}_{y}}{H_{2}} +
c_{1}{\bf r}_{x}+c_{2}{\bf r}_{t}, \quad {\bf e}_{3}={\bf 
e}_{1}\wedge{\bf e}_{2} \eqno(249)
$$
where $x=x^{1}, y=x^{2}, t=x^{3}$.
Now the equations  for ${\bf e}_{k}$ take the form (see, e.g., [59])
$$
\left ( \begin{array}{c}
{\bf e}_{1} \\
{\bf e}_{2} \\
{\bf e}_{3}
\end{array} \right)_{x}= B_{1}
\left ( \begin{array}{c}
{\bf e}_{1} \\
{\bf e}_{2} \\
{\bf e}_{3}
\end{array} \right), \quad
\left ( \begin{array}{c}
{\bf e}_{1} \\
{\bf e}_{2} \\
{\bf e}_{3}
\end{array} \right)_{y}= B_{2}
\left ( \begin{array}{c}
{\bf e}_{1} \\
{\bf e}_{2} \\
{\bf e}_{3}
\end{array} \right), \quad
\left ( \begin{array}{c}
{\bf e}_{1} \\
{\bf e}_{2} \\
{\bf e}_{3}
\end{array} \right)_{t}= B_{3}
\left ( \begin{array}{c}
{\bf e}_{1} \\
{\bf e}_{2} \\
{\bf e}_{3}
\end{array} \right)
\eqno(250)
$$
with
$$
B_{1}=
\left ( \begin{array}{ccc}
0             & k     &  -\sigma \\
-\beta k      & 0     & \tau  \\
\beta\sigma        & -\tau & 0
\end{array} \right) ,
B_{2}=
\left ( \begin{array}{ccc}
0            & m_{3}  & -m_{2} \\
-\beta m_{3} & 0      & m_{1} \\
\beta m_{2}  & -m_{1} & 0
\end{array} \right)
$$
$$
B_{3}=
\left ( \begin{array}{ccc}
0       & \omega_{3}  & -\omega_{2} \\
-\beta\omega_{3} & 0      & \omega_{1} \\
\beta\omega_{2}  & -\omega_{1} & 0
\end{array} \right) \eqno(251)
$$
where $k, \tau, \sigma, m_{i}, \omega_{i}$ are some real functions the 
explicit forms of which given in [55], $\beta={\bf e}_{1}^{2}=\pm 1, 
{\bf e}^{2}_{2}={\bf e}^{2}_{3}=1$. 
Again from the integrability condition of these equations 
we obtain the  set of equations 
$$
B_{1y}-B_{2x}+[B_{1},B_{2}]=0  \eqno(252a)
$$
$$
B_{1t}-B_{3x}+[B_{1},B_{3}]=0  \eqno(252b)
$$
$$
B_{3y}-B_{2t}+[B_{3},B_{2}]=0.  \eqno(252c)
$$

 Many 
integrable systems in 2+1 
dimensions are exact reductions of the equation (252) (see, e.g., [59]).

\subsection{Integrable reductions of 3-nonorthogonal coordinate systems. 
Examples}

Now to find out particular integrable 
reductions of 3-NOCSs, as in [54-57] ,
we will use MISSs. To 
this end, we assume that 
$$
{\bf e}_{1} \equiv {\bf S} \eqno(253)
$$
where ${\bf S}=(S_{1}, S_{2}, S_{3})$ is the spin vector, ${\bf 
S}^{2}=\beta=\pm 1$. So, the vector ${\bf e}_{1}$ satisfies the some 
given MISS.  Some comments on MISSs 
are in order. At present there exist several MISSs (see, e.g., 
[32,39]). They play important 
role both in mathematics and physics.  In this note, to find out a 
integrable 
case of 3-NOCSs,  we will use the several  
MISS. Before going into the problem, we would like to make the following 
observation. The metric $g_{ij}$ can be expressed by the spin vector ${\bf 
S}$. To show it, we put  
$$ {\bf r}_{x}^{2}=H^{2}_{1}=\beta =\pm 1.  
\eqno(254)
$$
Then we have
that
$$
{\bf r}=\partial^{-1}_{x}{\bf S}+{\bf r}_{0}(y,t)   \eqno(255)
$$
where $\partial^{-1}_{x}=\int^{x}_{-\infty}dx$. For simplicity,
we put ${\bf r}_{0}=0$. Now we
can express the coefficients of the metric (232) by ${\bf S}$. As shown in
[55], for the (2+1)-dimensional case, we have $$
g_{11} = {\bf S}^{2}=\beta=\pm 1, \quad g_{12}={\bf S}\cdot 
\partial^{-1}_{x}{\bf
S}_{y}, \quad  g_{13}={\bf S}\cdot\partial^{-1}_{x}{\bf S}_{t}
$$
$$
g_{22}=(\partial^{-1}_{x}{\bf S}_{y})^{2},
\quad g_{23}=(\partial^{-1}_{x}{\bf S}_{y})\cdot (\partial^{-1}_{x}{\bf
S}_{t}), \quad g_{33}=(\partial^{-1}_{x}{\bf S}_{t})^{2}.
\eqno(256)
$$

\subsubsection{The M-I equation}
One of the simplest MISSs is the Myrzakulov I (M-I) equation which reads as
[58]
$$
{\bf S}_t = ({\bf S} \wedge {\bf S}_{y}+u{\bf S})_{x}  \eqno(257a)
$$
$$
u_{x}=-{\bf S}({\bf S}_x \wedge {\bf S}_y).
\eqno(257b)
$$
This equation  is integrable by Inverse Scattering Transform (IST) 
(see, e.g., [58]). In this case,  we get ($\sigma=0$)
$$
m_{1}=\partial_{x}^{-1}[\tau_{y}+km_{2}), 
\quad m_{2}=\frac{u_{x}}{k}, \quad
m_{3}=\partial_{x}^{-1}(k_y -\tau m_{2})  \eqno(258)
$$
and 
$$
\omega_{1}= \frac{k_{xy}}{k}-\tau \partial^{-1}_{x}\tau_{y}, 
\quad
\omega_{2}=-k_{y},
\quad
\omega_{3} = 
-k\partial^{-1}_{x}\tau_{y}.
\eqno(259) $$
                                                                    
Thus we expressed the functions $m_{k}$ and $\omega_{k}$ by the three 
functions $k,\tau, \sigma$ and their derivatives. This means that we 
identified the equations (250) and (259) which define the geometry of 
three-dimensional flat  space with the given MISS - the equation (257). 

Now let us  we introduce two 
complex functions 
$q, p$ as 
$$ 
q = a_{1}e^{ib_{1}}, \quad p=a_{2}e^{ib_{2}}  \eqno(260)
$$
where $a_{j}, b_{j}$ are real functions. Let $a_{k}, b_{k}$ have the form
$$
a_{2} =\beta a_{1}=\beta\frac{k}{2}, \quad 
b_{1}=-b_{2}=-\partial^{-1}_{x}\tau.
 \eqno(261)
$$
It is easy to verify that the functions  $q,p$
satisfy the following (2+1)-dimensional NLS equation
$$
iq_t=q_{xy}+vq \eqno(262a)
$$
$$
-ip_t=p_{xy}+vp \eqno(262b)
$$
$$
v_{y}=2(pq)_{x} \eqno(262c)
$$
where $p=\beta \bar q$.
\subsubsection{The Ishimori  equation}
The Ishimori equation (IE) has the form [42]
$$
{\bf S}_t = {\bf S} \wedge ({\bf S}_{xx}+\alpha^{2} S_{yy})
+u_{y}{\bf S}_x+u_{x}{\bf S}_y  \eqno(263a)
$$
$$
u_{xx}-\alpha^{2}u_{yy}=-2\alpha^{2}{\bf S}({\bf S}_x \wedge {\bf S}_y).
\eqno(263b)
$$
The IE is also integrable by IST [42].
In this case,  we get
$$
m_{1}=\partial_{x}^{-1}[\tau_{y}-\frac{1}{2\alpha^2}M^{\prime}_2 u],\quad
m_{2}=-\frac{1}{2\alpha^2 k}M^{\prime}_2 u
$$
$$
m_{3}=\partial_{x}^{-1}[k_y +\frac{\tau}{2\alpha^2 k}M_2 u],
\quad M^{\prime}_{2}u=\alpha^{2}u_{yy}-u_{xx}  \eqno(264)
$$
and
$$
\omega_{2}= -(k_{x}+\sigma \tau)-
\alpha^{2}(m_{3y}+m_{2}m_{1})
+m_{2}u_{x}+\sigma u_{y}
$$
$$
\omega_{3}= (\sigma_{x}-k \tau)+
\alpha^{2}(m_{2y}-m_{3}m_{1})
+k u_{y}+m_{3}u_{x}
$$
$$
\omega_{1} = \frac{1}{k}[\sigma_{t}-\omega_{2x}+\tau\omega_{3}]. \eqno(265)
$$
Now let us  we consider two
complex functions
$q, p$ (260), where the explicit forms of $a_{j}, b_{j}$ given, for 
example, in [59]. 
Then the functions $q,p$ satisfy the L-equivalent and G-equivalent 
counterpart of the IE, namely, the Davey-Stewartson 
(DS) equation $$ iq_t+q_{xx}+\alpha^{2}q_{yy}+vq=0 \eqno(266a)
$$
$$
ip_t-p_{xx}-\alpha^{2}p_{yy}-vp=0 \eqno(266b)
$$
$$
\alpha^{2}v_{yy}--v_{xx}=-2[\alpha^{2}(pq)_{yy}+(pq)_{xx}]. \eqno(266c)
$$

\subsubsection{The case: $\sigma=\tau=m_{k}=\omega_{k}=0 (k=1,2)$}

Now we would like consider the simplest case when
$$
\sigma=\tau=m_{1}=m_{2}=\omega_{1}=\omega_{2}=0. \eqno(267)
$$
So the equation (252)
takes the form
$$
m_{3x}=k_{y}
\eqno(268a)
$$
$$
k_{t}=\omega_{3x}
\eqno(268b)
$$
$$
m_{3t}=\omega_{3y}.\eqno(268c)
$$
Let us now we consider the M-X equation [58]
$$
{\bf S}_{t}=\frac{\omega_{3}}{k}{\bf S}_{x} \eqno(269)
$$
with
$$
\omega_{3}=-k_{xx}-3k^{2}-3\alpha^{2}\partial^{-1}_{x}m_{3y}.  \eqno(270)
$$
With the $\omega_{3}$ of the form (270), the equation for $k$, as follows 
from (268),  takes the form
$$
k_{t}+6kk_{x}+k_{xxx}+3\alpha^{2}m_{3y}=0   \eqno(271a)
$$
$$
m_{3x}=k_{y} \eqno(271b)
$$
that is the nothing but the Kadomtsev-Petviashvili (KP)  equation.

\subsubsection{Diagonal metrics}

For the completeness of our exposition of the integrable cases of 
3-curvilinear coordinate systems, here we present the well-known fact [33]. 
Consider the case when the metric has the diagonal form, i.e. 
$$
ds^{2}=\epsilon_{1}H_{1}^{2}dx^{2} 
+\epsilon_{2}H^{2}_{2}dy^{2}+\epsilon_{3}H_{3}^{2}dt^{2}  \eqno(272) 
$$
where $\epsilon_{i}=\pm 1.$ In this note we consider the case when 
$\epsilon_{i}=+1$.  In this case, the Christoffel symbols take the form
$$
\Gamma^{k}_{ij}=0 
\quad i\neq 
j\neq k 
\eqno(273a)
$$
$$
\Gamma^{i}_{il}=\frac{H_{i,l}}{H_{i}}=\frac{H_{l}}{H_{i}}\beta_{li}  
\eqno(273b)
$$
$$
\Gamma^{i}_{ll}=-\frac{H_{l}H_{li}}{H_{i}^{2}}=-\frac{H_{l}}{H_{i}}\beta_{il}, 
\quad i\neq l. \eqno(273c) 
$$
Here
$$
\beta_{ik}=\frac{H_{k,i}}{H_{i}} \eqno(274)
$$
are the so-called rotation coefficients. 
      In this case we get 
$$
\tau=m_{2}=\omega_{3}=0  
\eqno(275) 
$$ 
and the matrices $B_{i}$ take the 
form 
$$
B_{1} =
\left ( \begin{array}{ccc}
0             & -\beta_{21}     &  -\beta_{31} \\
\beta_{21}       & 0     &   0 \\
\beta_{31}        &    0 & 0
\end{array} \right),\quad
B_{2}=
\left ( \begin{array}{ccc}
0            & \beta_{12}  & 0 \\
-\beta_{12} & 0      & -\beta_{32}\\
0  & \beta_{32} & 0
\end{array} \right)
$$
$$
B_{3} =
\left ( \begin{array}{ccc}
0       & 0  & \beta_{13} \\
0 & 0      & \beta_{23} \\
-\beta_{13}  & -\beta_{23} & 0
\end{array} \right).\eqno(276)
$$ 
So we have
$$
\frac{\partial \beta_{ij}}{\partial x^{k}}=\beta_{ik}\beta_{kj}, \quad 
i\neq j \neq k  \eqno(277a) $$
$$
\frac{\partial \beta_{ij}}{\partial 
x^{i}}+\frac{\partial \beta_{ji}}{\partial x^{j}}
+\sum^{3}_{m\neq i,j} \beta_{mi}\beta_{mj}=0, \quad i\neq j.  \eqno(277b)
$$                                      
This nonlinear system is the famous Lame equation and well-known in the 
theory of 3-orthogonal 
coordinates. The problem of description of curvilinear orthogonal 
coordinate systems in  a (pseudo-) Euclidean space is a classical 
problem of differential geometry. It was studied in detail and mainly 
solved in the beginning of the 20th century. Locally, such coordinate 
systems are determinated by $\frac{n(n-1)}{2}$ arbitrary functions of two 
variables.  This problem in 
some sense is equivalent to the problem of description of diagonal flat 
metrics, that is, flat metrics $g_{ij}(x)=f_{i}(x)\delta_{ij}$. 
We mention that the Lame equation describing curvilinear orthogonal 
coordinate systems can be  integrated by 
the IST [33] (see also an algebraic-geometric approach in [34]). 
                 
\subsection{Connections with the other equations}
It is remarkable that the equation (252) [=(318)=(316)=(294)] is related 
with the some well-known equations of modern theoretical physics. In this 
section we present some of these connections.

\subsubsection{ The Bogomolny equation}

Consider the Bogomolny 
equation (BE) [31] $$
\Phi_{t}+[\Phi,B_{3}]+ B_{1y} - B_{2x} + [B_{1},B_{2}] = 0 \eqno(278a)
$$
$$
\Phi_{y}+[\Phi,B_{2}]+B_{3x}-B_{1t}+[B_{3},B_{1}]=0   \eqno(278b)
$$
$$
\Phi_{x}+[\Phi,B_{1}]+B_{2t}-B_{3y}+[B_{2},B_{3}]=0. \eqno(278c)
$$
This equation is integrable and play important role in the field theories 
in particular in the theory of monopols. 
The set of equations (252) is the particular case of the BE. In fact, 
as $\Phi = 0$ from (278) we obtain the system (252). 

\subsubsection{The Self-Dual 
Yang-Mills  equation} 

Equation (252) is exact reduction of the SO(3)-SDYM equation
$$
F_{\alpha\beta}=0, \quad F_{\bar\alpha\bar\beta}=0,
\quad
F_{\alpha\bar\alpha}+F_{\beta\bar\beta}=0.  \eqno(279)
$$
Here
$$
F_{\mu\nu}=\frac{\partial A_{\nu}}{\partial x_{\mu}}-
\frac{\partial A_{\mu}}{\partial x_{\nu}}+[A_{\mu},A_{\nu}] \eqno(280)
$$
and
$$
\frac{\partial}{\partial x_{\alpha}}=\frac{\partial}{\partial z}-
i\frac{\partial}{ \partial t}, \quad
\frac{\partial}{\partial x_{\bar\alpha}}=\frac{\partial}{\partial z}+
i\frac{\partial}{ \partial t}, \quad
\frac{\partial}{\partial x_{\beta}}=\frac{\partial}{\partial x}-
i\frac{\partial}{ \partial y}
$$
$$
\frac{\partial}{\partial x_{\bar\beta}}=\frac{\partial}{\partial x}+
i\frac{\partial}{ \partial y}. \eqno(281)
$$
In fact, if in the SDYME (200) we take
$$
A_{\alpha}= -iB_{3}, \quad A_{\bar\alpha}= iB_{3}, \quad
A_{\beta}= B_{1}-iB_{2}, \quad A_{\beta}= B_{1}+iB_{2} \eqno(282)
$$
and if $B_{k}$ are independent of $z$, then the SDYM equation  (279)
reduces to  the  equation (252).
As known that the LR of the SDYM equation has the form [41, 43]
$$
(\partial_{\alpha}+\lambda\partial_{\bar\beta})\Psi=(A_{\alpha}+\lambda 
A_{\bar \beta})\Psi, 
\quad
(\partial_{\beta}-\lambda\partial_{\bar\alpha})\Psi=(A_{\beta}-
\lambda A_{\bar\alpha})\Psi \eqno(2283)
$$
where $\lambda$ is the spectral parameter  satisfing the following
set of the equations
$$
\lambda_{\beta}=\lambda\lambda_{\bar\alpha}, \quad
\lambda_{\alpha}=-\lambda\lambda_{\bar\beta}. \eqno(284)
$$
Apropos, the simplest solution of this set has may be the following
form 
$$
\lambda 
=\frac{a_{1}x_{\bar\alpha}+a_{2}x_{\bar\beta}+a_{3}}{a_{2}x_{\alpha}-
a_{1}x_{\beta}+a_{4}}, \quad a_{j}=consts. \eqno(285)
$$
From (284) we obtain the LR of the equation (252)
$$
(-i\partial_{t}+\lambda\partial_{\bar\beta})\Psi=[-iB_{3}+\lambda 
(B_{1}+iB_{2})]\Psi \eqno(286a)
$$
$$
(\partial_{\beta}-i\lambda\partial_{t})\Psi=[(B_{1}-iB_{2})-i\lambda 
B_{3}]\Psi.
 \eqno(286b)
$$     

\subsubsection{The Chern-Simons equation}

Consider the action of the Chern-Simons (CS) theory [44]
$$
S[J]=\frac{k}{4\pi}\int_{M}tr(J\wedge dJ+\frac{2}{3}J\wedge J\wedge J)
\eqno(287)
$$
where $J$ is a 1-form gauge connection with values in the Lie algebra 
$\hat g$ of a (compact or noncompact) non-Abelian simple Lie group $\hat 
G$ on an oriented closed 3-dimensional manifold $M$, $k$ is the coupling 
constant. The classical equation of motion is the zero-curvature condition
$$ 
dJ+J\wedge J=0.  \eqno(288) $$ 
Let the 1-form $J$ has the form
$$
J=B_{1}dx+B_{2}dy+B_{3}dt.  \eqno(289)
$$
As shown in [44], subtituting the (289) into (288) we obtain the equation 
(252).  Note that from this fact and from the results of 
the subsection 5.2 follows that the CS - equation of motion (288) is exact 
reduction of the SDYM equation (279).

\subsubsection{The Lame equation}

Let us the matrices $B_{i}$ (251) we rewrite in the form
$$
B_{1} =
\left ( \begin{array}{ccc}
0             & -\beta_{21}     &  -\beta_{31} \\
\beta_{21}       & 0     &   \tau \\
\beta_{31}        & -\tau & 0
\end{array} \right),\quad
B_{2}=
\left ( \begin{array}{ccc}
0            & \beta_{12}  & -m_{2} \\
-\beta_{12} & 0      & -\beta_{32}\\
m_{2}  & \beta_{32} & 0
\end{array} \right)
$$
$$
B_{3} =
\left ( \begin{array}{ccc}
0       & \omega_{3}  & \beta_{13} \\
-\omega_{3} & 0      & \beta_{23} \\
-\beta_{13}  & -\beta_{23} & 0
\end{array} \right).\eqno(290)
$$

Then the equation (252) in elements takes the form
$$
\beta_{23x}-\tau_{t}=\beta_{13}\beta_{21}-\omega_{3}\beta_{31} \eqno(291a)
$$
$$
 \beta_{32x}+\tau_{y}=\beta_{12}\beta_{31}+m_{2}\beta_{21}
   \eqno(291b)
$$
$$
\beta_{13y}+m_{2t}=\beta_{12}\beta_{23}+\omega_{3}\beta_{32}
\eqno(291c)
$$
$$
 \beta_{31y}-m_{2x}=\beta_{32}\beta_{21}-\tau\beta_{12}
   \eqno(291d)
$$
$$
\beta_{12t}-\omega_{3y}=\beta_{13}\beta_{32}-m_{3}\beta_{23}
\eqno(291e)
$$
$$
 \beta_{21t}+\omega_{3x}=\beta_{23}\beta_{31}+\tau\beta_{13}   \eqno(291f)
$$
$$
\beta_{12x}+\beta_{21y}+\beta_{31}\beta_{32}+\tau m_{2}=0
\eqno(291g)
$$
$$
\beta_{13x}+\beta_{31t}+\beta_{21}\beta_{23}+\tau\omega_{3}=0
   \eqno(291h)
$$
$$
\beta_{23y}+\beta_{32t}+\beta_{12}\beta_{13}+m_{2}\omega_{3}=0.
   \eqno(291i)
$$
Hence as 
$\tau=m_{2}=\omega_{3}=0 $
 we obtain the Lame equation (277). So, the equation (252) is one of the 
generalizations of the Lame equation.

\subsection{On Lax representations in multidimensions}

As follows from the results of  the  previous subsection, 
Equation (252) can  
admits several integrable reductions. At the same time, the results of 
the subsections 3.2 show that the equation (252) is integrable 
may be and in general case.  At least, it admits the LR of the form (286) 
and/or of the following form (see, e.g., [38]) 
$$
\Phi_{x}=U_{1}\Phi, \quad  \Phi_{y}=U_{2}\Phi, \quad \Phi_{t}=U_{3}\Phi 
\eqno(292) 
$$
with
$$
U_{1}=\frac{1}{2} 
\left ( \begin{array}{cc}
i\tau          & -(k+i\sigma) \\
k-i\sigma       & -i\tau
\end{array} \right),\quad
U_{2}=
\frac{1}{2}\left ( \begin{array}{cc}
im_{1}       & -(m_{3}+im_{2}) \\
m_{3}-im_{2} &  -im_{1}
\end{array} \right)
$$
$$
U_{3}=
\frac{1}{2}\left ( \begin{array}{cc}
i\omega_{1}       & -(\omega_{3}+i\omega_{2}) \\
\omega_{3}-i\omega_{2} &  -i\omega_{1}
\end{array} \right).  \eqno(293)
$$
Systems of this type were first studied by Zakharov and Shabat [35]. The 
integrability conditions on this system of overdetermined equations 
(292), require that 
$$
U_{i,j}-U_{j,i}+[U_{i}, U_{j}]=0. \eqno(294)
$$
Many (and perhaps all) integrable systems in 2+1 dimensions have the LR of 
the form (292). 
In our case, e.g., the IE (263) and the DS equation (266) have also the 
LR of the form (292) with 
the functions $m_{i}, \omega_{i}$ given by (264) and (265). On the other 
hand, it is well-known that  for example  the DS equation (266) has the 
following LR of the standard form
$$
\alpha\Psi_{y}=\sigma_{3}\Psi_{x}+Q\Psi,  \quad 
Q=\left ( \begin{array}{cc}
0   & q\\
p   &  0
\end{array} \right)  
 \eqno(295a)
$$
$$
\Psi_{t}=2i\sigma_{3}\Psi_{xx}+2iQ\Psi_{x}+
\left ( \begin{array}{cc}
c_{11}       & iq_{x}+i\alpha q_{y} \\
ip_{x}-i\alpha p_{y} &  c_{22}
\end{array} \right)\Psi  \eqno(295b)
$$
with
$$
c_{11x}-\alpha c_{11y}=i[(pq)_{x}+\alpha(pq)_{y}], \quad 
c_{22x}+\alpha c_{22y}=-i[(pq)_{x}-\alpha(pq)_{y}]. \eqno(296)
$$

Hence arises the natural question: how connected the 
both LR for one and the same integrable systems (in our case 
for the DS equation)?.                    
In fact, these two LR are related by the gauge transformation [54-56]
$$
\Phi=g \Psi  \eqno(297)
$$
where $\Phi$ and $\Psi$ are some solutions of the equations (292) and (295), 
respectively, while $g$ is the some matrix.

\section{The n=4 case} Now we wish consider the physically interesting
case when $n=4$.  In this case the equation (246) takes the form 
$$ 
\left (
\begin{array}{c} {\bf e}_{1} \\ {\bf e}_{2} \\ {\bf e}_{3}\\ {\bf e}_{4}
\end{array} \right)_{x}= B_{1} \left ( \begin{array}{c} {\bf e}_{1} \\
{\bf e}_{2} \\ {\bf e}_{3}\\ {\bf e}_{4} \end{array} \right), \quad \left
( \begin{array}{c} {\bf e}_{1} \\ {\bf e}_{2} \\ {\bf e}_{3}\\ {\bf e}_{4}
\end{array} \right)_{y}= B_{2} \left ( \begin{array}{c} {\bf e}_{1} \\
{\bf e}_{2} \\ {\bf e}_{3}\\ {\bf e}_{4} \end{array} \right) $$ $$ \left (
\begin{array}{c} {\bf e}_{1} \\ {\bf e}_{2} \\ {\bf e}_{3}\\ {\bf e}_{4}
\end{array} \right)_{z}= B_{3} \left ( \begin{array}{c} {\bf e}_{1} \\
{\bf e}_{2} \\ {\bf e}_{3}\\ {\bf e}_{4} \end{array} \right), \quad \left
( \begin{array}{c} {\bf e}_{1} \\ {\bf e}_{2} \\ {\bf e}_{3}\\ {\bf e}_{4}
\end{array} \right)_{t}= B_{4} \left ( \begin{array}{c} {\bf e}_{1} \\
{\bf e}_{2} \\ {\bf e}_{3}\\ {\bf e}_{4} \end{array} \right) \eqno(298) $$
where the matrices $B_{k}$ belong already to the Lie algebra $so(p,q)$,
i.e., $B_{k}\in so(p,q), \quad p+q=4.$ Hence we have $$
B_{1y}-B_{2x}+[B_{1},B_{2}]=0 \eqno(299a) 
$$ 
$$ 
B_{1z}-B_{3x}+[B_{1},B_{3}]=0
\eqno(299b) 
$$ 
$$ 
B_{1t}-B_{4x}+[B_{1},B_{4}]=0 \eqno(299c) 
$$ 
$$
B_{2z}-B_{3y}+[B_{2},B_{3}]=0 \eqno(299d) 
$$ 
$$ B_{2t}-B_{4y}+[B_{2},B_{4}]=0
\eqno(299e) 
$$ 
$$ 
B_{3t}-B_{4z}+[B_{3},B_{4}]=0. \eqno(299f)
$$

It is often convenient to reformulate 
equations (298) in terms of 
the matrices. To do so, we introduce the matrices $\hat e_{k}$ in 
such a way that the determinant of which is equal to 1:
$$
\hat e_{k}=
\left ( \begin{array}{cc}
e_{k4}+\epsilon_{3}e_{k3}& -(\epsilon_{1}e_{k1}+i\epsilon_{2}e_{k2}) \\
\epsilon_{1}e_{k1}-i\epsilon_{2}e_{k2} &  e_{k4}-\epsilon_{3}e_{k3}
\end{array} \right).  \eqno(300)
$$
Hence and from the definition of the vectors ${\bf e}_{k}$ follows that 
$$
det(\hat e_{k})=\epsilon^{2}_{1}e_{k1}^{2}+\epsilon^{2}_{2}e^{2}_{k2}+ 
\epsilon_{3}^{2} e_{k3}^{2}+e_{k4}^{2}=1 \eqno(301)
$$
where $\epsilon_{k}^{2}=\pm 1$. 
The matrices $\hat e_{k}$ satisfy the set of 16 equations
$$
\hat e_{i,j}=b^{j}_{ik}\hat e_{k} \eqno(302)
$$
where $b^{j}_{ik}$ are the elements of $B_{j}$.
\subsection{Integrable 
reductions of 4-nonorthogonal coordinate systems.
Example} In contrast with the $n=3$ case, for the $n=4$ we have only a 
few number integrable systems in 3+1 dimensions such as the generalized 
and intrinsic generalized wave and sine-Gordon equations [46-47] (for the 
case $n=4$) and so on. Among them the  
SDYM equation takes the special place. We need in the following form, 
e.g., of the anti-SDYM  equation (see, e.g., [49])
$$
(g^{-1}g_{z_{1}})_{\bar z_{1}}+(g^{-1}g_{z_{2}})_{\bar z_{2}}=0 \eqno(303)
$$
where $g\in SL(2,C), \quad \det g =1, \quad  g>0$ and
$$
z_{1}=\frac{1}{2}(x_{4}+ix_{3}), \quad 
z_{2}=\frac{1}{2}(x_{2}+ix_{1}).
\eqno(304) 
$$ 
Equation (303) is integrable by IST. The corresponding LR has the form [?]
$$
(\lambda \partial_{z_{1}}-\partial_{\bar 
z_{2}}+\lambda g^{-1}g_{z_{1}})\Phi=0 \eqno(305a)
$$
$$
( \partial_{\bar z_{1}}+\lambda\partial_{
z_{2}}+\lambda g^{-1}g_{z_{2}})\Phi=0. \eqno(305b)
$$

{\bf Conjecture 1} [56]. {\it If the 
matrix} $\hat e_{1}$ (300) {\it satisfies the anti-SDYM equation (303), 
i.e.,} $$
(\hat e_{1}^{-1}\hat e_{1z_{1}})_{\bar 
z_{1}}+(e_{1}^{-1}e_{1z_{2}})_{\bar z_{2}}=0 \eqno(306)  
$$
{\it then the equation (299) for the case $n=4$ with the matrices $B_{k}\in 
so(p,q), \quad p+q=4$ is integrable}.

\section{On soliton immersions in multidimensions}
In this section we would like discuss some aspects of simplest soliton 
immersions 
in multidimensions (see also [46-48]). The results in the study of this 
problem in the Fokas-Gelfand sense [1,3] will be reported elsewhere.

\subsection{$V^{3}$ in $R^{\mu}$}
Let $R^{\mu}, \quad \mu=(p,q), \quad p+q=4$ be the 4-dimensional  space
 with 
the coordinate system $(y)=(y^{k})$ and the metric
$$
ds^{2}=\mu_{ij}dy^{i}dy^{j}.   \eqno(307)
$$          

Let $V^{3}$ is 3-dimensional submanifolds of the $R^{\mu}$ with the 
coordinates $(x)=(x^{k})$ and the metric
$$
ds^{2}=g_{ij}dx^{i}dx^{j}.   \eqno(308)
$$

Let $F: \Omega \rightarrow R^{\mu}$ be an immersion of a domain $\Omega\in 
V^{3}$ into $R^{\mu}$. Let $(x)=(x^{1}=x, x^{2}=y, x^{3}=t)\in \Omega$. 
Let ${\bf n}(x,y,t)$ be the normal vector field defined at each point of 
the three-dimensional submanifolds  $F(x,y,t)$. The space $F(x,y,t)$ is 
defined by the first (308) and second fundamental forms
$$
II=-dF\cdot {\bf n}= L_{ij}dx^{i}dx^{j}.   \eqno(309)
$$
 
Then the system of vectors $(F_{x}, F_{y}, F_{t}, n)$ defines a basis of 
$V^{3}$ on $S$ parametrized by $F(x,y,t)$.
Let ${\bf 
r}=(y^{1},y^{2},y^{3},y^{4}) 
={\bf r}(x^{1},x^{2},x^{3})$ is the position vector. 
Then the position vector ${\bf r}$ 
satisfies the following Gauss-Weingarten (GW) equation
$$
{\bf r}_{xx} = \Gamma^{1}_{11} {\bf r}_{x} + \Gamma^{2}_{11} {\bf r}_{y}+ 
\Gamma^{3}_{11} {\bf r}_{t}+L_{11}{\bf n}   \eqno(310a) $$
$$
{\bf r}_{xy} = \Gamma^{1}_{12} {\bf r}_{x} + \Gamma^{2}_{12} {\bf r}_{y}+ 
\Gamma^{3}_{12} {\bf r}_{t} + L_{12}{\bf n} \eqno(310b)
$$
$$
{\bf r}_{xt} = \Gamma^{1}_{13} {\bf r}_{x} + 
\Gamma^{2}_{13} {\bf r}_{y}+ \Gamma^{3}_{13} {\bf r}_{t}+L_{13}{\bf n}  
\eqno(310c) $$
$$
{\bf r}_{yy} = \Gamma^{1}_{22} {\bf r}_{x} + \Gamma^{2}_{22} {\bf r}_{y}+ 
\Gamma^{3}_{22} {\bf r}_{t}+L_{22}{\bf n}  \eqno(310d)
$$
$$
{\bf r}_{yt} = \Gamma^{1}_{23} {\bf r}_{x} + \Gamma^{2}_{23} {\bf r}_{y}+ 
\Gamma^{3}_{23} {\bf r}_{t}+L_{23}{\bf n}  \eqno(310e)
$$
$$
{\bf r}_{tt} = \Gamma^{1}_{33} {\bf r}_{x} + \Gamma^{2}_{33} {\bf r}_{y}+ 
\Gamma^{3}_{33} {\bf r}_{t}+L_{33}{\bf n}  \eqno(310f)
$$
$$
{\bf n}_{x} = p_{11} {\bf r}_{x} + p_{12} {\bf r}_{y}+
p_{13} {\bf r}_{t}   \eqno(310g) 
$$
$$
{\bf n}_{y} = p_{21} {\bf r}_{x} + p_{22} {\bf r}_{y}+
p_{23} {\bf r}_{t}   \eqno(310h)
$$   
$$
{\bf n}_{t} = p_{31} {\bf r}_{x} + p_{32} {\bf r}_{y}+
p_{33} {\bf r}_{t}.   \eqno(310i)
$$   
Here $\Gamma^{k}_{ij}$ are the Cristoffel's symbols defined by
$$
\Gamma^{k}_{ij}=\frac{1}{2}g^{kl}(g_{il,j}+g_{jl,i}-g_{ij,l})   \eqno(311)
$$
and $$
L_{ij}=({\bf r}_{,ij}\cdot {\bf n}).  \eqno(312)
$$

We can rewrite the GW equation (310) in the following form
$$
Z_{x} = A_{1}Z, \quad Z_{y} = A_{2}Z, \quad Z_{t} = 
A_{3}Z  \eqno(313) $$
where
$$
Z=({\bf r}_{x}, {\bf r}_{y}, {\bf r}_{t}, {\bf n})^{T}
\eqno(314)
$$ 
and
$$
A_{1} =
\left ( \begin{array}{cccc}
\Gamma^{1}_{11} & \Gamma^{2}_{11} & \Gamma_{11}^{3}&L_{11} \\
\Gamma^{1}_{12} & \Gamma^{2}_{12} & \Gamma^{3}_{12}&L_{12} \\
\Gamma^{1}_{13} & \Gamma^{2}_{13} & \Gamma^{3}_{13}&L_{13} \\ 
p_{11}           & p_{12}           & p_{13}       &0
\end{array} \right) ,  \quad
A_{2} =
\left ( \begin{array}{cccc}
\Gamma^{1}_{11} & \Gamma^{2}_{11} & \Gamma_{11}^{3}&L_{11} \\
\Gamma^{1}_{12} & \Gamma^{2}_{12} & \Gamma^{3}_{12}&L_{12} \\
\Gamma^{1}_{13} & \Gamma^{2}_{13} & \Gamma^{3}_{13}&L_{13} \\
p_{11}           & p_{12}           & p_{13}       &0
\end{array} \right) ,  
$$
$$
A_{3} =
\left ( \begin{array}{cccc}
\Gamma^{1}_{11} & \Gamma^{2}_{11} & \Gamma_{11}^{3}&L_{11} \\
\Gamma^{1}_{12} & \Gamma^{2}_{12} & \Gamma^{3}_{12}&L_{12} \\
\Gamma^{1}_{13} & \Gamma^{2}_{13} & \Gamma^{3}_{13}&L_{13} \\
p_{11}           & p_{12}           & p_{13}       &0
\end{array} \right). 
\eqno(315)
$$
Integrability conditions for GW equations
$$
\frac{\partial A_{i}}{\partial x^{j}}-\frac{\partial A_{j}}{\partial 
x^{i}}  + [A_{i},A_{j}] = 0  \eqno(316)
$$
are known as Gauss-Mainardi-Codazzi (GMC) equations.

\subsubsection{Integrable reductions}

First let us rewrite the equation (313) in terms of the triad ${\bf 
e}_{k}$. We have
$$
\left ( \begin{array}{c}
{\bf e}_{1} \\
{\bf e}_{2} \\
{\bf e}_{3}\\
{\bf e}_{4}
\end{array} \right)_{x}= B_{1}
\left ( \begin{array}{c}
{\bf e}_{1} \\
{\bf e}_{2} \\
{\bf e}_{3}\\
{\bf e}_{4}
\end{array} \right)
\eqno(317a)
$$
$$
\left ( \begin{array}{c}
{\bf e}_{1} \\
{\bf e}_{2} \\
{\bf e}_{3}\\
{\bf e}_{4}
\end{array} \right)_{y}= B_{2}
\left ( \begin{array}{c}
{\bf e}_{1} \\
{\bf e}_{2} \\
{\bf e}_{3}\\
{\bf e}_{4}
\end{array} \right)
\eqno(317b)
$$
$$
\left ( \begin{array}{c}
{\bf e}_{1} \\
{\bf e}_{2} \\
{\bf e}_{3}\\
{\bf e}_{4}
\end{array} \right)_{t}= B_{3}
\left ( \begin{array}{c}
{\bf e}_{1} \\
{\bf e}_{2} \\
{\bf e}_{3}\\
{\bf e}_{4}
\end{array} \right).
\eqno(317c)
$$
Here $B_{k}\in so(p,q), \quad p+q=4$.  The compatibility condition of the 
linear equations (317) is
$$
B_{1y}-B_{2x}+[B_{1}, B_{2}]=0 \eqno(318a)
$$
$$
B_{1t}-B_{3x}+[B_{1}, B_{3}]=0 \eqno(318b)
$$
$$
B_{2t}-B_{3y}+[B_{2}, B_{3}]=0. \eqno(318c)
$$

In terms of the matrices the 
equation (317) takes the form $$
\hat e_{jx}=b^{1}_{ki}\hat e_{i}
\eqno(319a)
$$
$$
\hat e_{jy}=b^{2}_{ki}\hat e_{i}
\eqno(319b)
$$
$$
\hat e_{jt}=b^{3}_{ki}\hat e_{i}
\eqno(319c)
$$
where $j=1,2,3 \quad \hat e_{k}$ are given by (300),  $b^{j}_{ki}$ are 
elements of $B_{j}$. Now we give two examples of integrable cases. 
For the other examples and for details, see, e.g., [55-56].
\\
\\
\\
i)  {\bf  The Manakov-Zakharov-Mikhailov equation}
\\
\\

Consider the Manakov-Zakharov-Mikhailov (MZM) equation
[50-51]
$$
(g^{-1}g_{t})_{t}-(g^{-1}g_{\xi})_{\eta}=0 \eqno(320)
$$
where $g$ is some $2\times 2$  matrix and
$$
 \det g =1, \quad \xi=\frac{1}{2}(x+\sigma y), \quad 
\eta=\frac{1}{2}(x-\sigma y), \quad  \sigma^{2}=\pm 1. \eqno(321)
$$
The MZM equation is equivalent to the compatibility condition for the two 
linear problems
$$
(\lambda 
\partial_{\eta}-\lambda^{-1}\partial_{\xi}-\lambda^{-1}g^{-1}g_{\xi}+ 
g^{-1}g_{t})\Phi=0 \eqno(322a) 
$$
$$
( \partial_{t}+\lambda\partial_{\eta}+g^{-1}g_{t})\Phi=0 \eqno(322b)
$$
or [49]
$$
(\partial_{\xi}+\lambda\partial_{t}+g^{-1}g_{\xi})\Phi=0 \eqno(323a) 
$$
$$
(\lambda\partial_{\eta}+\partial_{t}+g^{-1}g_{t})\Phi=0. \eqno(323b)
$$

{\bf Conjecture 2} [55-56]. {\it If the
matrix} $\hat e_{1}$ (300) {\it satisfies the MZM equation (320), i.e.,}
$$
(\hat e^{-1}_{1}\hat e_{1t})_{t}-(\hat e_{1}^{-1}\hat e_{1\xi})_{\eta}=0 
\eqno(324) 
$$
{\it then the equation (318) for the case $n=4$ with the matrices $B_{k}\in
so(p,q), \quad p+q=4$ is integrable}.
\\
\\
\\
ii)  {\bf The generalized sigma model}
\\
\\

As next example, we consider the following generalized sigma model [52]
$$
(\Sigma_{y})_{x} 
+(\Sigma_{t})_{t}+\alpha[(\Sigma_{t})_{x}-(\Sigma_{x})_{t}]=0. \eqno(325) 
$$
This equation is integrated by the LR
[52]
$$
[\partial_{y}+(\alpha+k)\partial_{t}+(\Sigma_{y}+\alpha\Sigma_{t})]
\Phi=0 \eqno(326a)
$$
$$
[-(\alpha+k) 
\partial_{x}+\partial_{t}+(-\alpha\Sigma_{x}+\Sigma_{t})]
\Phi=0 
\eqno(326b) $$
where $\Sigma_{i}=g^{-1}\partial_{i}g$. Hence as $\alpha=0$, we get the 
Ward model [53].

{\bf Conjecture 3}. {\it If the
matrix $\hat e_{1}$ (300) satisfies the generalized sigma model (325), i.e.,}
$$
\Sigma_{i}=\hat e^{-1}_{1}\partial_{i}\hat e_{1}
\eqno(327)
$$
{\it then the equation (318) for the case $n=4$ with the matrices $B_{k}\in
so(p,q), \quad p+q=4$ is integrable}.

\subsection{$V^{4}$ in $R^{\mu}$}
Similarly, we can consider the manifold $V^{4}$ embedded in $R^{\mu}$ 
where $\mu=(p,q)$ with $p+q=n>4$. 
 In this case the equations for ${\bf e}_{k}$ take the form 
$$ 
\left (\begin{array}{c} 
{\bf e}_{1} \\ 
{\bf e}_{2} \\ 
... \\
{\bf e}_{n}
\end{array} \right)_{,k}= B_{k}
 \left ( \begin{array}{c} {\bf e}_{1} \\
{\bf e}_{2} \\ 
... \\
{\bf e}_{n}
\end{array} \right)
\eqno(328)
$$
where the matrices $B_{k}$ belong already to the Lie algebra $so(p,q)$,
i.e., $B_{k}\in so(p,q), \quad p+q=n.$ The matrices $B_{k}$  satisfy the 
system
$$
B_{1y}-B_{2x}+[B_{1},B_{2}]=0 \eqno(329a) 
$$ 
$$ 
B_{1z}-B_{3x}+[B_{1},B_{3}]=0
\eqno(329b) 
$$ 
$$ 
B_{1t}-B_{4x}+[B_{1},B_{4}]=0 \eqno(329c) 
$$ 
$$
B_{2z}-B_{3y}+[B_{2},B_{3}]=0 \eqno(329d) 
$$ 
$$ 
B_{2t}-B_{4y}+[B_{2},B_{4}]=0
\eqno(329e) 
$$ 
$$ 
B_{3t}-B_{4z}+[B_{3},B_{4}]=0. \eqno(329f)
$$

{\bf Conjecture 4} [56]. {\it If the
matrix} $\hat e_{1}$ (300) {\it satisfies the anti-SDYM equation (303), 
i.e.,} $$
(\hat e_{1}^{-1}\hat e_{1z_{1}})_{\bar
z_{1}}+(e_{1}^{-1}e_{1z_{2}})_{\bar z_{2}}=0 \eqno(330)  
$$
{\it then the equation (329) for the case $n=4$ with the matrices $B_{k}\in
so(p,q), \quad p+q=4$ is integrable}.

\section{Conclusion}
We see that the geometrical equations, describing $n$-NOCSs in 
flatEuclidian and pseudo-Euclidian spaces, admit several integrable 
reductions.  Moreover, we have conjectured that the immersions of 3-, and 
4-dimensional manifolds arbitrarily embedded in $R^{\mu}$ admit 
integrable cases.  

Concluding  this work we would like to note that  as it 
seems we still very 
far from the understanding the geometrical nature of multidimensional 
integrable 
systems. So this subject needs further developments and making more 
precise. We shall end here.

\section{Acknowledgments}
This work was partially supported by INTAS (grant 99-1782). RM would like 
to thanks to V.Dryuma, M.Gurses, B.Konopelchenko, D.Levi, L.Martina and 
G.Soliani for very helpful discussions and especially  M.Gurses and 
D.Levi for the financial supports and kind hospitality. He is grateful to 
the EINSTEIN Consortium of Lecce University 
and the Department of Mathematics of Bilkent University for their
financial supports and warm hospitality.

\end{document}